\documentclass[aps,pra,showpacs,twocolumn,superscriptaddress]{revtex4}
\usepackage{graphicx,color}
\usepackage{amsmath,amsthm,amsfonts,amssymb,bm}
\usepackage{times}
\usepackage{epsf}
\usepackage[colorlinks={true}]{hyperref}

\usepackage[T1]{fontenc}
\usepackage[utf8]{inputenc}

\hypersetup{citecolor={blue}, filecolor={blue}, linkcolor={blue}, urlcolor={blue}}

\newcommand{\commentold}[1]{}
\DeclareMathSymbol{:}{\mathpunct}{operators}{"3A}

\begin{document}

\title{Enhancement of Frequency Estimation by Spatially Correlated Environments}
\author{R. Yousefjani}
\author{S. Salimi}
\email{shsalimi@uok.ac.ir}
\author{A. S. Khorashad}

\affiliation{Department of Physics, University of Kurdistan, P.O.Box 66177-15175 , Sanandaj, Iran}

\date{\today}

\begin{abstract}

In metrological tasks, employing entanglement can quantitatively improve the precision of parameter estimation.
However, susceptibility of the entanglement to decoherence fades this capability in the realistic metrology and limits ultimate quantum improvement.
One of the most destructive decoherence-type noise is uncorrelated Markovian noise which commutes with the parameter-encoding Hamiltonian and is modelled as a semigroup dynamics, for which the quantum improvement is constrained to a constant factor.
It has been shown [Phys. Rev. Lett. \textbf{109}, 233601 (2012)] that when the noisy time evolution is governed by a local and non-semigroup dynamics (e.g., induced by an uncorrelated non-Markovian dephasing), emerging the Zeno regime at short times can result in the Zeno scaling in the precision.
Here, by considering the impact of the correlated noise in metrology, we show that spatially correlated environments which lead to a nonlocal and non-semigroup dynamics can improve the precision of a noisy frequency measurement beyond the Zeno scaling.
In particular, it is demonstrated that one can find decoherence-free subspaces and subsequently achieve the Heisenberg precision scaling for an approximated dynamics induced by spatially correlated environments.

\end{abstract}


\maketitle

\section{Introduction}\label{I}
Quantum metrology is rooted in the quantum estimation theory and aims at achieving the best possible precision in estimating quantities which cannot be analyzed by direct measurement, such as phase, frequency, or magnetic fields.
Due to the existence of such quantities in all branches of physics like gravitational-wave detection \cite{LIGO}, atomic clocks \cite{Roos,Ospelkaus}, frequency spectroscopy \cite{Wineland,Bollinger} and interferometry \cite{Resch,Higgins}, quantum metrology has emerged as an active field of research.

The great advantage of quantum metrology relies on the effective use of entanglement which allows one to attain precision that is beyond the ability of classical metrology.
While the lowest estimation uncertainty in classical protocols with $N$  probes is the standard quantum limit ($\mathcal{O}(N^{-1/2})$) \cite{Pezz,Itano}, quantum protocols with entangled probes can lead to the Heisenberg limit ($\mathcal{O}(N^{-1})$) \cite{Leibfried,Roos,Giovannetti}.
Although taking advantage of entanglement in quantum protocols improves the precision of estimation, it makes the protocoles very sensitive to noise.
Hence, it is important to investigate the restrictions which are dictated on the ultimate achievable precision by the dynamics arising from the presence of an environment \cite{Demk,Escher,Maccone}.
According to the scale of correlation times during which correlations disappear, noises can be grouped into Markovian and non-Markovian types.
Subject to an uncorrelated (local) Markovian noise whose dynamics is ruled by a semigroup, an arbitrarily small noise level is enough to limit the ultimate precision to the standard quantum limit \cite{Huelga,Demk}.
However, it has been shown \cite{Jan} that when such Markovian noise is spatially correlated (nonlocal) one can identify decoherence-free subspaces and obtain the Heisenberg precision scaling.
Regarding an uncorrelated non-Markovian dephasing, Matsuzaki et al. \cite{Matsuzaki} and Chin et al. \cite{Chin}, have shown that non-semigroup dynamics assists the maximal entangled probes to provide the Zeno scaling of $N^{-3/4}$ order.
This improvement is a direct consequence of the quadratic time dependence of the dynamics (the Zeno dynamics) at short times.
Recently, the generality of the Zeno scaling has been argued for any frequency estimation in the presence of local phase covariant noise \cite{K,Demk2}.

This paper focuses on the impacts of the correlated (nonlocal) non-semigroup dynamics on the estimation uncertainty.
Actually, initial correlations {among} environments can generate a nonlocal dynamics in an open quantum system, even if system-environment interactions are local \cite{Elsi,Steffen,Liu}.
In this case, the system exhibits features which are not present in the dynamics of the individual subsystems.
{By deriving the noise factor as the probability with which the environmental state makes a transition from its initial state to any other state, we show that correlated environments make much slower transitions than those made by uncorrelated ones.}
It should be emphasized that {such} transitions have various temporal behaviors at short times.
By relating this phenomenon to the correlation of the environmental coupling operators, we {consider} the necessary and sufficient conditions for the
occurrence of transitions which are governed by $t^{4}$.
Such transitions allow the entangled probes to experience a longer evolution which decreases the estimation uncertainty scaling to $\mathcal{O}(N^{-7/8})$.
In addition, it is demonstrated that spatially correlated non-semigroup noise can also provide the ability to determine decoherence-free subspaces and obtain the Heisenberg scaling if one ignores the free evolution of the environments during the interactions.
To show these, an organized Ramsey spectroscopy setup is considered in the presence of the pure dephasing.

The structure of the paper is as follows.
In section $\textbf{II}$, physical model, we design our Ramsey spectroscopy setup in the presence of the pure dephasing.
In section $\textbf{III}$, after calculation of coherence factor as the result of the environmental transitions, the influences of the initial correlations on its temporal behavior are studied.
The possibility of the existence of the decoherence-free subspaces in the presence of the correlated noise and subsequently achieving the Heisenberg precision scaling are discussed in section $\textbf{IV}$.
This paper concludes in section $\textbf{V}$.

\section{Physical model}\label{III}

Let us consider a scheme consisting of $N$ identical probes. Every single probe, $\textit{S}^{\mu}$ $ $ $(\mu=1,..,N)$, in turn, comprises two two-level atoms (labeled by an index  $i=1,2$) with different transition frequencies $\nu_{1}$ and $\nu_{2}$. The frequency difference $\bar{\nu}=\nu_{1}-\nu_{2}$ is going to be estimated by performing standard Ramsey-type measurement. To generate the Ramsey pulses and derive atomic transition from $|0\rangle$ to $|1\rangle$, two lasers with frequencies $\nu_{L1}$ and $\nu_{L2}$ are utilized.
Hence, all of the atoms with frequency $\nu_{1}$ accumulate relative phase $\varepsilon_{1}=\nu_{1}-\nu_{L1}$ and the others with frequency $\nu_{2}$ pick up $\varepsilon_{2}=\nu_{2}-\nu_{L2}$.
The uncertainty in the estimation of $\bar{\nu}$ is given by the square root of the statistical average of the squared differences between the true and the estimated values of the parameter, $(\delta\bar{\nu})$. The lowest of this uncertainty is given by the Cramér–Rao bound as \cite{Caves2,Caves, Book}
\begin{equation}\label{Cramer-Rao bound}
(\delta\bar{\nu})^{2}=\frac{1}{\frac{T}{t}\mathcal{F}(\bar{\nu})},
\end{equation}
\begin{figure}[b]
\includegraphics[scale=.56]{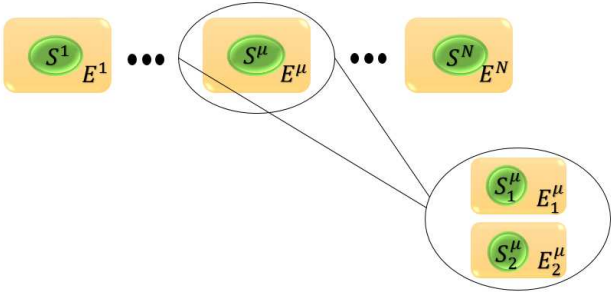}
\caption{(Color online) Schematic illustrating the local interaction of any probe with the associated environment. The inset gives the probe-environment interaction in detail.   }
\label{fig1}
\end{figure}
where $t$ is the duration of each single measurement, $T/t$ is the number of times that one can perform the measurement for a given fixed $T$, and $\mathcal{F}(\bar{\nu})$ is the Fisher information. Considering $p$ as the probability of coincidence of the initial and final states at time $t$, $\mathcal{F}(\bar{\nu})$ can be evaluated as
\begin{equation}\label{Fisher information}
\mathcal{F}(\bar{\nu})=\frac{| \partial p/\partial \bar{\nu} |^{2}}{p(1-p)}.
\end{equation}
The $N$ probes suffer from the associated $N$ independent environments $\textit{E}^{\mu}$ $ $ $(\mu=1,..,N)$ which induce pure decoherence (See Fig. \ref{fig1}).
{This kind of noise is often the most important factor limiting the attainable precision in spectroscopy.}
Every environment $\textit{E}^{\mu}$ is formed of two multimode bosonic subenvironments which locally interact with the corresponding qubits of each probe (See inset of Fig. \ref{fig1}).
Performing $N$ independent measurements on the $N$ uncorrelated probes leads to the Fisher information $\mathcal{F}_{N}(\bar{\nu})$ which is $N$ times greater than the Fisher information of a single probe, $\mathcal{F}_{1}(\bar{\nu})$, and therefore the uncertainty is given as
\begin{equation}
(\delta\bar{\nu})^{2}=\frac{1}{N}\frac{1}{\frac{T}{t}\mathcal{F}_{1}(\bar{\nu})}.
\end{equation}
Suppose that the initial state of a single probe is $|\Psi_{12}\rangle=\frac{1}{\sqrt{2}}(|\textbf{\textsf{0}}\rangle+|\textbf{\textsf{1}}\rangle)$ $ $ $(|\textbf{\textsf{0}}\rangle=|01\rangle$ and $|\textbf{\textsf{1}}\rangle=|10\rangle)$. After an evolution period of time $t$ in the presence of pure dephasing, the reduced state in the Schr\"{o}dinger picture is obtained as (see Appendix)
\begin{eqnarray}\label{Roh}
\nonumber \rho(t) &=& \frac{1}{2}\Bigg(|\textbf{\textsf{0}}\rangle\langle\textbf{\textsf{0}}| +|\textbf{\textsf{1}}\rangle\langle\textbf{\textsf{1}}| \\
\nonumber &+& e^{i\bar{\varepsilon} t+\gamma(t)}|\textbf{\textsf{0}}\rangle\langle\textbf{\textsf{1}}|+
e^{-i\bar{\varepsilon} t+\gamma(t)}|\textbf{\textsf{1}}\rangle\langle\textbf{\textsf{0}}|\Bigg),\\
\end{eqnarray}
where $\bar{\varepsilon}=\varepsilon_{1}-\varepsilon_{2}$ is the relative phase and $\gamma(t)=\gamma_{0110}(t)$ is the decay rate.
After some calculations, the corresponding expression of Eq. (3) is obtained as
\begin{equation}\label{Cramer-Rao bound for uncorrelated probes}
(\delta\bar{\nu})^{2}=\frac{1}{N}\frac{1-e^{2\gamma (t)}cos^{2}(\bar{\varepsilon} t)}{T t e^{2\gamma (t)}sin^{2}(\bar{\varepsilon} t)}.
\end{equation}
It can be proved \cite{Chin} that Eq. (5) attains its minimum value if the relative phase is chosen to be $\bar{\varepsilon}=k \pi/2t_{u}$ (for odd $k$) and the optimal duration of each single measurement (optimal interrogation time), $t_{u}$ ($u$ refers to uncorrelated probes), satisfies
\begin{equation}\label{optiamal t uncorrelated}
1+2t\partial_{t}\gamma(t)\mid_{t=t_{\textit{u}}}=0.
\end{equation}
Using these conditions, one has
\begin{equation}\label{best resolution for uncorrelated probes}
(\delta\bar{\nu})^{2}\mid_{\textit{u}}=\frac{1}{N}\frac{1}{Tt_{\textit{u}}e^{2\gamma (t_{\textit{u}})}}.
\end{equation}
For an initial preparation of $N$ probes in maximally entangled state
$\frac{1}{\sqrt{2}}(|\textbf{\textsf{0}}\rangle^{\otimes N}+|\textbf{\textsf{1}}\rangle^{\otimes N})$, the analogous calculations for independent queries subject to the noise originated from $N$ independent environments lead to
\begin{equation}\label{best resolution for correlated probes}
(\delta\bar{\nu})^{2}\mid_{\textit{e}}=\frac{1}{N^{2}}\frac{1}{Tt_{\textit{e}}e^{2N\gamma (t_{\textit{e}})}},
\end{equation}
where $t_{\textit{e}}$ ($e$ refers to entangled probes) satisfies
\begin{equation}\label{optiamal t correlated}
1+2Nt\partial_{t}\gamma(t)\mid_{t=t_{\textit{e}}}=0.
\end{equation}
It is obvious that $\gamma(t)$ is the key factor to specify the optimal interrogation time and subsequently the scaling of the lowest uncertainty.
This factor is the result of transitions induced in the environment $\textit{E}^{\mu}$ by the system-environment interaction.
It can be seen \cite{Chin,Demk2} that the time dependence of such quantum mechanical transitions at short time after the system-environment interaction is switched on is an effective factor in determining the scale of $t_{\textit{e}}$.
In the following, after calculating the general form of $\gamma(t)$ for Gaussian environments, the time dependence of such transitions under various conditions is investigated.

%%%%%%%%%%%%%%%%%%%%%%%%%%%%%%%%%%%%%%%%%%%%%%%%%%%%%%%%%%%%%%%%%%%%%%%%%%%%%%%%%%%%%%%%%%%%%%%%%%%%%%%%%%%%%%%%%%%%%%%%%%%%%%%%%%%%%%%%%%%%%%%%%%%%%%%%%%%%%%%%%
%%%%%%%%%%%%%%%%%%%%%%%%%%%%%%%%%%%%%%%%%%%%%%%%%%%%%%%%%%%%%%%%%%%%%%%%%%%%%%%%%%%%%%%%%%%%%%%%%%%%%%%%%%%%%%%%%%%%%%%%%%%%%%%%%%%%%%%%%%%%%%%%%%%%%%%%%%%%%%%%%

\section{calculation of $\gamma(t)$}
As is shown in the Appendix, regarding the environmental initial state as a tensor product of two-mode states, $\rho_{E}(0)=\bigotimes_{l}\varrho_{l}^{12}$, the decay rate can be obtained as $\gamma(t)=\gamma_{0110}(t)=\Sigma_{l}\ln[\mathcal{X}_{l}^{1001}]$, where
$\mathcal{X}_{l}^{1001}=Tr[\mathcal{D}[-2\beta_{l}^{(1)}]\otimes \mathcal{D}[2\beta_{l}^{(2)}]\varrho_{l}^{12} ]$ is the Wigner characteristic function of the state $\varrho_{l}^{12}$, $\mathcal{D}$ is the displacement operator, $\beta_{l}^{(i)}=g_{l}^{(i)}(1-e^{i\omega_{l}^{(i)}t})/\omega_{l}^{(i)}$ and the coefficient $g_{l}^{(i)}$ shows the coupling strength between the $i$th subsystem and its own environment with mode $l$.
Here, the environmental states are chosen from the Gaussian states family because they have an experimental realization and simple mathematical structure.
In general, a Gaussian state is completely characterized by the first and second statistical moments of the quadrature field operators which in natural units, $\hbar =2$, are defined as  \cite{Adesso1,Adesso2}
\begin{eqnarray}
 \nonumber q_{l} &=& (b_{l}^{\dagger}+b_{l}) \\
  p_{l} &=& (b_{l}^{\dagger}-b_{l})/i ,
\end{eqnarray}
where $b_{l}$ and $b_{l}^{\dagger}$ are annihilation and creation (bosonic field) operators of the mode $l$, respectively.
Without loss of generality one may assume that the Gaussian state has zero mean.
Hence, the second moments form the covariance matrix $\mathbf{\sigma}=[\sigma_{ij}=\frac{1}{2}\langle\{R_{i},R_{j}\}\rangle_{\varrho_{l}^{12}}]$ where in the case of a two-mode Gaussian state one has $R=(q_{1},p_{1},q_{2},p_{2})^{T}$.
The diagonal elements of $\mathbf{\sigma}$ are directly linked to the mean energy of any mode but the off-diagonal ones denote the quadrature correlations  (quantum and classical) which vanish for a product state.
There exist local symplectic operations which can transfer a two-mode covariance matrix into a standard form as \cite{Duan}
\begin{equation}\label{covariance matrix}
 \mathbf{\sigma} = \begin{pmatrix} a & 0 & c_{+} & 0\\ 0 & a & 0 & c_{-}\\ c_{+} & 0 & b & 0 \\ 0 & c_{-} & 0 & b \end{pmatrix},
\end{equation}
{where $a=\langle q_{1}^{2}\rangle $, $b=\langle q_{2}^{2}\rangle $, $c_{+}=\langle q_{1}q_{2}\rangle$ and $c_{-}=\langle p_{1}p_{2}\rangle$.
Note that $\det(\sigma)+1\geq a^{2}+b^{2}+2c_{+}c_{-}$.
This condition certifies
that $\varrho_{l}^{12}\geq 0$ holds, so it is a physical state \cite{Simon}.}
{Here, }$\varrho_{l}^{12}$ is a symmetric state ($a=b$) due to the uniformity of the subenvironments. One can assume, without loss of generality, that $c_{+}\geq0$ and $c_{-}=\theta c_{+}$ $(-1\leq\theta\leq 1)$.
\\Using this covariance matrix the decay factor is given as
\begin{eqnarray}\label{total decay factor1}
 \nonumber \gamma(t) &=& \sum_{l}\Bigg(-8a \Bigg(|\beta_{l}^{(1)}|^{2}+|\beta_{l}^{(2)}|^{2}\Bigg) \\
\nonumber   &-& 4c_{+}(1-\theta)  \Bigg(\beta_{l}^{(1)}\beta_{l}^{(2)}+\beta_{l}^{(1)\ast}\beta_{l}^{(2)\ast} \Bigg) \\
   &+& 4c_{+}(1+\theta)\Bigg(\beta_{l}^{(1)}\beta_{l}^{(2)\ast}+\beta_{l}^{(1)\ast}\beta_{l}^{(1)}\Bigg) \Bigg).
\end{eqnarray}
After performing the continuum limit with the same Ohmic spectral density, $J=\omega \exp[-\omega/\omega_{c}]$, for both multimode bosonic subenvironments  $\textit{E}^{\mu} _{1}$ and $\textit{E}^{\mu} _{2}$,  the decay factor can be simplified to
\begin{eqnarray}\label{decay factor2}
\nonumber \gamma (t) &=& -8(a-\theta c_{+})\ln[1+\omega_{c}^{2}t^{2}] \\
   &+&2c_{+}(1-\theta)\ln[1+4\omega_{c}^{2}t^{2}].
\end{eqnarray}
%%%%%%%%%%%%%%%%%%%%%%%%%%%%%%%%%%%%%%%%%%%%%%%%%%%%%%%%%%%%%%%%%%%%%%%%%%%%%%%%%%%%%%%%%%%%%%%%%%%%%%%%%%%%%%%%%%%%%%%%%%%%%%%%%%%%%%%%%%%%%%%%%%%%%%%%%
%%%%%%%%%%%%%%%%%%%%%%%%%%%%%%%%%%%%%%%%%%%%%%%%%%%%%%%%%%%%%%%%%%%%%%%%%%%%%%%%%%%%%%%%%%%%%%%%%%%%%%%%%%%%%%%%%%%%%%%%%%%%%%%%%%%%%%%%%%%%%%%%%%%%%%%%%
%%%%%%%%%%%%%%%%%%%%%%%%%%%%%%%%%%%%%%%%%%%%%%%%%%%%%%%%%%%%%%%%%%%%%%%%%%%%%%%%%%%%%%%%%%%%%%%%%%%%%%%%%%%%%%%%%%%%%%%%%%%%%%%%%%%%%%%%%%%%%%%%%%%%%%%%%
\subsection{Uncorrelated noise }
In the absence of the initial correlations between the two subenvironments $\textit{E}^{\mu} _{1}$ and $\textit{E}^{\mu} _{2}$ (i.e., $c_{+}=c_{-}=0$) inducing local process, the decay factor has a quadratic behavior, $\gamma^{\textit{local}}(t)\approx -8a\omega_{c}^{2}t^{2}$, at times much shorter than correlation time of the noise ($\omega_{c}^{-1}$). This universal time dependence results in the optimal interrogation time $(t_{\textit{e}}^{\textit{local}})$ being obtained from Eq. (\ref{optiamal t correlated}) as
\begin{equation}\label{q for local}
 t_{\textit{e}}^{\textit{local}}\cong\frac{\omega_{c}^{-1}}{\sqrt{32 N a}}.
\end{equation}
Therefore, non-semigroup dynamics enables the entangled probes to beat the standard quantum limit by providing   $(\delta\bar{\nu})\mid_{\textit{e}}^{\textit{local}}\propto N^{-3/4}$ which is the same Zeno scaling obtained by the other authors
\cite{Matsuzaki,Chin,K,Demk2}.
%%%%%%%%%%%%%%%%%%%%%%%%%%%%%%%%%%%%%%%%%%%%%%%%%%%%%%%%%%%%%%%%%%%%%%%%%%%%%%%%%%%%%%%%%%%%%%%%%%%%%%%%%%%%%%%%%%%%%%%%%%%%%%%%%%%%%%%%%%%%%%%%%%%%%%%%%
%%%%%%%%%%%%%%%%%%%%%%%%%%%%%%%%%%%%%%%%%%%%%%%%%%%%%%%%%%%%%%%%%%%%%%%%%%%%%%%%%%%%%%%%%%%%%%%%%%%%%%%%%%%%%%%%%%%%%%%%%%%%%%%%%%%%%%%%%%%%%%%%%%%%%%%%%
%%%%%%%%%%%%%%%%%%%%%%%%%%%%%%%%%%%%%%%%%%%%%%%%%%%%%%%%%%%%%%%%%%%%%%%%%%%%%%%%%%%%%%%%%%%%%%%%%%%%%%%%%%%%%%%%%%%%%%%%%%%%%%%%%%%%%%%%%%%%%%%%%%%%%%%%%
\subsection{Correlated noise }
Expanding Eq. (\ref{decay factor2}) up to the leading order in $\omega_{c}t$ gives
\begin{eqnarray}\label{Gamma1}
\nonumber   \gamma^{nonlocal}(t) &\approx& -8a(1-c_{+}/a)\omega_{c}^{2}t^{2} \\
   &+& 4a(1-(4-3\theta)c_{+}/a)\omega_{c}^{4}t^{4}.
\end{eqnarray}
{Note that for symmetric Gaussian states, the quotient $c_{+}/a = \langle q_{1}q_{2}\rangle/\sqrt{\langle q_{1}^{2}\rangle\langle q_{2}^{2}\rangle}$
is correlation coefficient which quantifies the correlation among the quadratures $q_{1}$ and $q_{2}$.}
It gives a value between $0$ and $1$, where $1$ denotes that the positions of the particles are strongly correlated and $0$ means no correlation.
From the structure of $ \gamma^{nonlocal}(t)$, one infers that although spatial correlation slows down the transitions in the environments, it cannot change their quadratic time dependence provided that it is not strong enough.

\begin{figure}
    \centering
    {
        \includegraphics[width=3in]{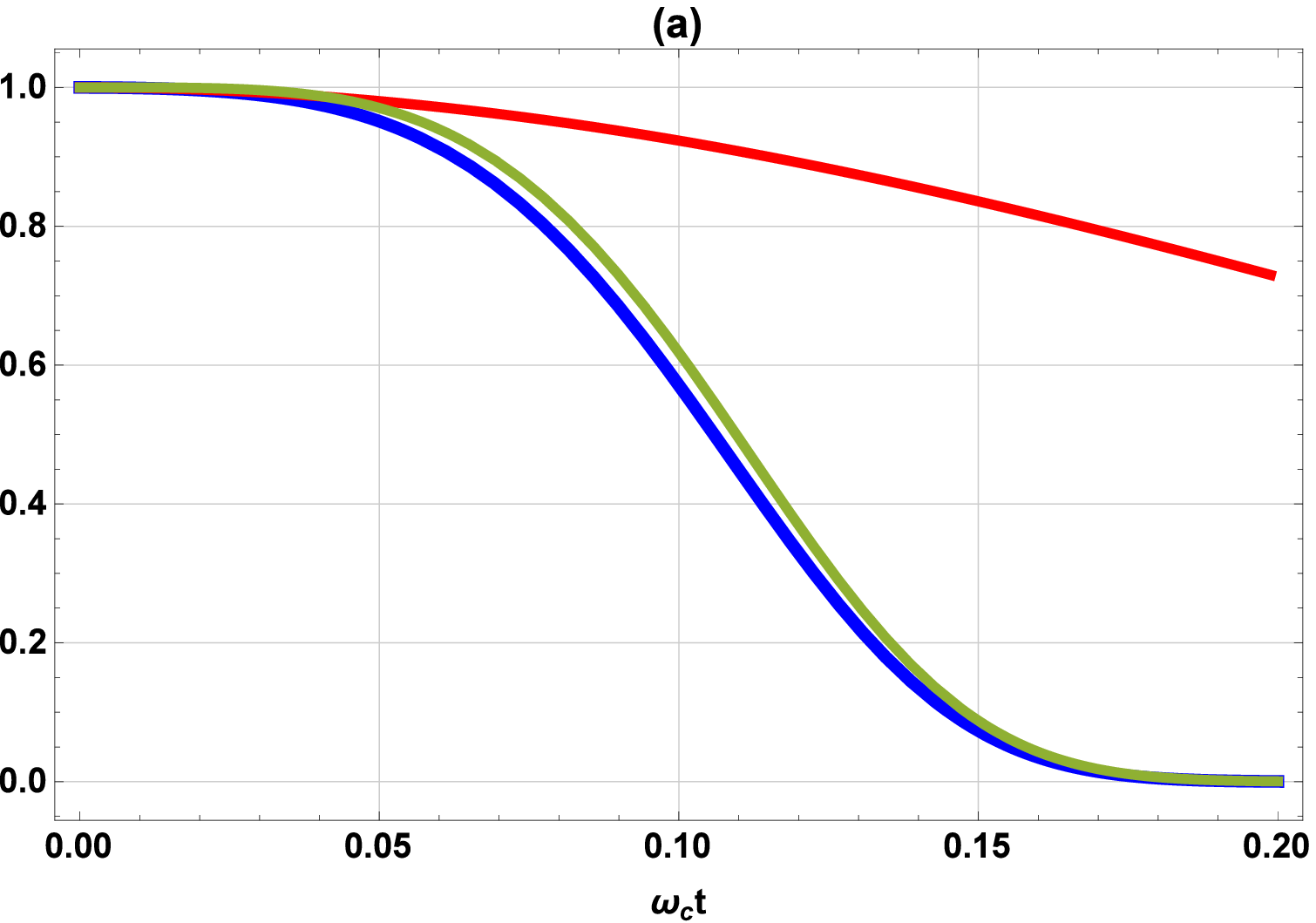}
    }
    \\
    {
        \includegraphics[width=3in]{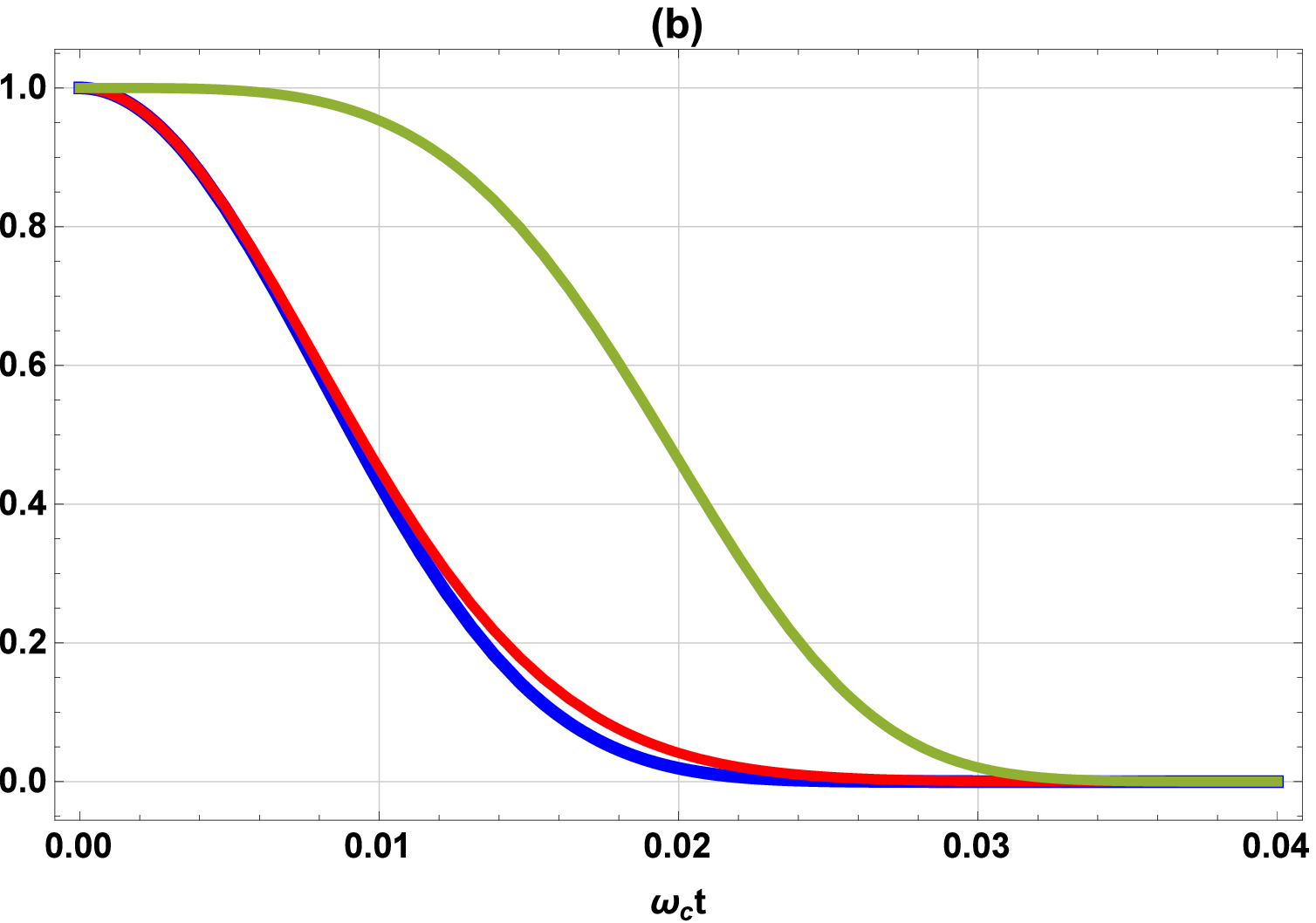}
    }

    \caption{(Color online) Dynamics of the coherence factor, $e^{2N\gamma^{nonlocal}(t)}$ (blue), $e^{2N(-8a(1-c_{+}/a)\omega_{c}^{2}t^{2})}$ (red) and $e^{2N(4a(1-7c_{+}/a)\omega_{c}^{4}t^{4})}$ (green) for correlated environments with $a\simeq 10$, $c_{+}/a \simeq .995$ and $\theta=-1$, using $N=10$ in (a) and $N\rightarrow \infty$ in (b). }
    \label{Fig2}
\end{figure}

{Regarding the number of particles, $N$, in the scheme of Fig. \ref{fig1}, two scenarios can be considered.
For finite $N$, strong correlation in particle positions, $c_{+}/a \rightarrow 1$, is the necessary condition for occurring transitions which are not governed by $t^{2}$.
In the asymptotic $N$ limit scenario, for strong correlation, but $c_{+}/a$ not exactly $1$, the term proportional to $t^{2}$ in Eq. (\ref{Gamma1}), albeit very small, will fix the asymptotic scaling of $N^{-3/4}$ for the ultimate precision.
So, in this case, just provided that $c_{+}/a = 1$, one can observe transitions in the environments which are governed by $t^{4}$.
To clarify these issues, we compare contributions of two components of $\gamma^{nonlocal}(t)$ in Fig. \ref{Fig2} for schemes which use (a) finite  and (b) infinite number of particles.
Since creation and protection of large amounts of entanglement are beyond the current technology, considering the case of finite $N$ seems more logical, as we do in the following.}

Depending on the values of $\theta$, the strong correlated environments can be grouped into two types; $(1)$ those with $\theta \neq 1$ and $(2)$ those with  $\theta = 1$.
The first class includes  all the environments with $q$- and $p$- quadratures which are unequally correlated, $c_{+}\neq c_{-}$.
Such environments satisfy the sufficient condition ($\theta \neq 1$) to observe transitions which are governed by $t^{4}$ at short times.
An important example of this class is pure two-mode Gaussian state with $\theta=-1$ and $c_{+}=\sqrt{a^{2}-1}$ which is strongly correlated for $a\gg 1$.
Any pure two-mode Gaussian state is actually equivalent to a two-mode squeezed vacuum state of the form \cite{Adesso1,Adesso2,Souza}
\begin{equation}\label{EPR-state}
|\eta_{l}^{12}\rangle=\sqrt{1-\tanh^{2}(r)}\sum_{n=0}^{\infty}\tanh^{n}(r)|n\rangle_{1}\otimes|n\rangle_{2},
\end{equation}
where $|n\rangle_{i}$ denotes the $n$th Fock state of the $i$th subenvironment and $r\in \mathbb{R}^{+}$ is the squeezing parameter.
It should be mentioned that for sufficiently large $r$ the state in Eq. (\ref{EPR-state}) can be considered as a good approximation for the Einstein–Podolsky–Rosen (EPR) state which is the analogue of a maximally entangled state for continuous variable systems.
Since for this state $a=\cosh(2r)$, the condition of strong correlation is satisfied by large $r$.

\begin{figure}[b]
\includegraphics[width=3in]{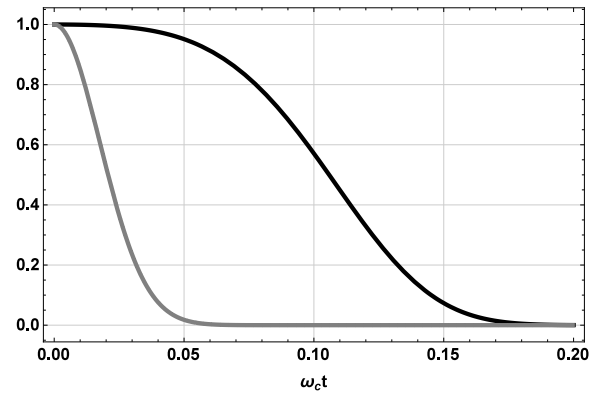}
\caption{(Color online) Dynamics of the coherence factor, $e^{\gamma(t)}$ for uncorrelated environments with $a=\cosh(3)$ and $c_{+}=c_{-}=0$ (black carve) and entangled environments with $a=\cosh(3)$ and $c_{+}=-c_{-}=\sinh(3)$ (Gray carve).       }
\label{fig3}
\end{figure}

The dynamics of the coherence factor, $e^{\gamma(t)}$, induced by this correlated (with $\theta\neq 1$) and an uncorrelated Gaussian states is plotted in Fig. \ref{fig3}.
The key point illustrated in this figure is that spatially correlated environments make slower transitions compared to those made by uncorrelated ones with a probability proportional to $t^{4}$.
From Eq. (\ref{optiamal t correlated}), this characteristic behavior results in an optimal interrogation time (for the entangled probes) scaling as
$\omega_{c}^{-1}N^{-1/4}$  which indicates that obtaining the optimal estimation requires a longer interrogating range.
The consequence of such time scale is $(\delta\bar{\nu})\mid_{\textit{e}}^{\textit{nonlocal}}\propto N^{-7/8}$ which shows that nonlocality of the non-semigroup dynamics can significantly enhance the precision of the estimation.

{As one can see from Eq. (\ref{decay factor2}), the strongly correlated environments, $c_{+}/a \cong 1$, with $\theta=1$ at times much shorter than the correlation time of the noise, result in $\gamma(t)\sim 0$.
So, the noise does not approximately affect the initially maximally entangled state.}
This shows that the initial probe's state, $\frac{1}{\sqrt{2}}(|\textbf{\textsf{0}}\rangle^{\otimes N}+|\textbf{\textsf{1}}\rangle^{\otimes N})$,
is in a decoherence-free subspace with respect to the noise.
Hence, frequency measurements in the presence of such special noise can be done by the Heisenberg precision scaling.
Possibility of the existence of the decoherence-free subspaces in the presence of the correlated noise and subsequently achieving the Heisenberg scaling is not limited to the environments with $\theta=1$.
In the next section, to prove this claim an approximated dynamics is introduced.
%%%%%%%%%%%%%%%%%%%%%%%%%%%%%%%%%%%%%%%%%%%%%%%%%%%%%%%%%%%%%%%%%%%%%%%%%%%%%%%%%%%%%%%%%%%%%%%%%%%%%%%%%%%%%%%%%%%%%%%%%%%%%%%%%%%%%%%%%%%%%%%%%%%%%%%%%%%%%%
%%%%%%%%%%%%%%%%%%%%%%%%%%%%%%%%%%%%%%%%%%%%%%%%%%%%%%%%%%%%%%%%%%%%%%%%%%%%%%%%%%%%%%%%%%%%%%%%%%%%%%%%%%%%%%%%%%%%%%%%%%%%%%%%%%%%%%%%%%%%%%%%%%%%%%%%%%%%%%
%%%%%%%%%%%%%%%%%%%%%%%%%%%%%%%%%%%%%%%%%%%%%%%%%%%%%%%%%%%%%%%%%%%%%%%%%%%%%%%%%%%%%%%%%%%%%%%%%%%%%%%%%%%%%%%%%%%%%%%%%%%%%%%%%%%%%%%%%%%%%%%%%%%%%%%%%%%%%%
\section{Decoherence-free subspaces}
It can be shown \cite{Jan} that in perfect correlated Markovian noise it is always possible to identify subspaces which are decoherence-free. Therefore,  the Heisenberg precision scaling in the frequency measurements becomes attainable by  the initial states which are chosen from these subspaces.
On the other hand, in the presence of nonlocal non-semigroup dymanics, the Heisenberg precision scaling can also be obtained by means of an approximated dynamics where the free evolution of each environment modes is neglected.
To obtain the time propagator for this dynamics, one can follow the procedure in the Appendix.
The result is $\tilde{U}_{i}(t)=\sum_{n_{i}=0}^{1}P_{n_{i}}\otimes u_{n_{i}}(t)$, where $u_{n_{i}}(t)=e^{-itB_{n_{i}}}$ which can be rewritten as  $u_{n_{i}}(t)=\otimes_{l}\mathcal{D}[(-1)^{n_{i}+1}\beta_{l}^{(i)}]$ with $\beta^{(i)}_{l}=-itg^{(i)}_{l}$.
Inserting $\beta^{(i)}_{l}$ in Eq. (\ref{total decay factor1}) results in $\gamma^{\textit{nonlocal}}(t)=-8a(1-c_{+}/a)\omega_{c}^{2}t^{2}$
if an Ohmic spectral density with the cutoff frequency $\omega_{c}$ is assumed.
Therefore, provided that $c_{+}/a \cong 1$, the initial preparation of $N$ probes in maximally entangled state,
$\frac{1}{\sqrt{2}}(|\textbf{\textsf{0}}\rangle^{\otimes N}+|\textbf{\textsf{1}}\rangle^{\otimes N})$, can lead to $(\delta\bar{\nu})\mid_{\textit{e}}^{\textit{nonlocal}}\propto N^{-1}$.
{Since the diminishing effect of the free evolution is almost negligible for very short interaction times, our conjecture is that at
$t \ll \omega_{c}^{-1}$,
strong correlation among the positions of the particles, $c_{+}/a \cong 1$, is the sufficient condition for the occurrence of the decoherence-free evolution.
}

\section{Results And Discussion} \label{VI}
Exploiting non-semigroup dynamics can lead to improved precision scaling. This is a direct consequence of nonlinear temporal behavior of the dynamics at short times.
The universal time dependence of induced transitions in uncorrelated environments, due to the system-environment interaction, is like $t^{2}$ (emerging the Zeno regime for open system dynamics) leading to quadratic decay of the coherence.
In this case, one finds out that interrogating the entangled probes in time intervals which scale as $N^{-1/2}$ enables the estimation uncertainty to be of the Zeno scaling, $\mathcal{O}(N^{-3/4})$.
By considering the impacts of the nonlocal non-semigroup dynamics (induced by correlated environments) in metrology, we have shown that spatial correlation in environments can improve precision scaling beyond the Zeno scaling{ in non-asymptotic limit}.
In general, the correlated environments have much sluggish transitions than those made by the uncorrelated ones which can lead to various decay behaviors.
This fact has been shown by considering an organized Ramsey spectroscopy in the presence of pure dephasing.
We have analyzed the dynamics of bosonic environments with the initial correlations and derived the necessary and sufficient conditions to observe transitions with the probabilities varying as $t^{4}$ at short times.
These environments authorize the entangled probes to have a longer evolution (scaling as $N^{-1/4}$) and to obtain more information about the parameter.
This indicates that in the presence of nonlocal non-semigroup dynamics {the entangled states of the systems with finite particles number} can decrease the uncertainty scaling to $\mathcal{O}(N^{-7/8})$.
Moreover, we have shown that there are special correlated environments without any diminishing effects on the coherence which allow to obtain the Heisenberg precision scaling in our frequency measurements.
It should be mentioned that without strong initial (spetial) correlation, it is not possible to get such results.
Finally, we have {clarified} that the spatially correlations among the environments with an approximated dynamics provide us with the ability to identify the decoherence-free subspaces and subsequently obtain the Heisenberg precision scaling in our frequency measurements.

{Decoherence induced by correlated environments is a commonly observed effect in biological light-harvesting complexes \cite{biologicla}, multi-atom trapping experiments \cite{eperiments with trapped ions}, photonic systems \cite{Liu} and etc.. Generally, there are different ways of modelling correlated noise and the approach with several baths which might have correlations, as has been discussed here, is a possible one.}
{Potential systems to which our model can be applied could include photons and ions.
Quantum optics experiments provides facilities for testing fundamental aspects of quantum mechanics.
In fact, photonic systems are one of the ideal candidates for the physical realization of quantum tasks due to the extremely high level of control over their degrees of freedom.
In all optical setups, controlled interactions between different degrees of freedom of a photon can be generated and the initial environmental states
can be selectively prepared.
In addition, in optical setups entanglement is an accessible resource and can be produced in modern optics labs with spontaneous parametric downconversion (SPDC).
A natural realization of dephasing noise in optics experiments can be achieved by coupling the polarization degree of freedom of a photon as a two-level system with its frequency (mode) degree of freedom as its environment.
There exists an experimental work \cite{Liu} on entangled photons with arbitrary polarization states and controllable correlation among the photon frequencies (environments).}
{Beyond optics experiments, spatially correlated environments can be generated by fluctuating electromagnetic fields in trapped and laser-cooled ions.
If the generated fluctuations by the field in one site can at least partially  be felt at adjacent site, then the fluctuations at  various sites are no longer independent and spatial correlations have to be taken into account.
Recent experiments with trapped ions located in close proximity to each other have proven to be dominated by spatially correlated dephasing \cite{eperiments with trapped ions}.}

\subparagraph{\textbf{Acknowledgments:}}
We would like to thank J. Jeske and A. T. Rezakhani for useful discussions.\\

\appendix{\textbf{Appendix:}}
Consider a system $\textit{S}$ which is composed of two two-level atoms labeled by an index  $i=1,2$. Each of the atoms interacts locally with its own multimode bosonic environment $\textit{E}_{i}$ where the total Hamiltonian is given as
\begin{equation}\label{total Hamiltonian}
\nonumber H=\sum_{i=1}^{2}H_{\textit{S}}^{(i)} + H_{\textit{E}}^{(i)}+H_{\textit{SE}}^{(i)},
\end{equation}
in which
\begin{eqnarray}\label{S and E Hamiltonian}
 \nonumber H_{\textit{S}}^{(i)} &=& \frac{\varepsilon_{i}}{2}\sigma_{z}^{(i)}, \\
 \nonumber  H_{\textit{E}}^{(i)} &=& \sum_{l}\omega_{l}^{(i)}b^{(i)\dag}_{l}b^{(i)}_{l},
\end{eqnarray}
and
\begin{equation}\label{SE Hamiltonian}
 \nonumber H_{\textit{SE}}^{(i)} = \sum_{l}\sigma_{z}^{(i)}\otimes\Bigg(g_{l}^{(i)}b^{(i)\dag}_{l}+g_{l}^{(i)*}b^{(i)}_{l}\Bigg),
\end{equation}
where $b^{(i)}_{l}$ and $b^{(i)\dagger}_{l}$ represent the annihilation and creation operators, respectively, of the $l$th mode of the $i$th environment. They satisfy commutation relation $[b^{(i)}_{l},b^{(j)\dag}_{l^{\prime}}]=\delta_{ij}\delta_{l l^{\prime}}$. The coefficient $g_{l}^{(i)}$ shows the coupling strength between the $i$th subsystem and its own environment with mode $l$.
By rewriting the system Hamiltonian as
\begin{eqnarray}\label{S and E Hamiltonian}
 \nonumber H_{\textit{S}}^{(i)} &=& \frac{\varepsilon_{i}}{2}\sum_{n_{i}=0}^{1}(-1)^{n_{i}+1}P_{n_{i}} \hspace{0.4cm} (P_{n_{i}}=|n_{i}\rangle\langle n_{i}|),
\end{eqnarray}
the interaction Hamiltonian can be presented as
\begin{equation}\label{SE Hamiltonian}
 \nonumber H_{\textit{SE}}^{(i)} = \sum_{n_{i}=0}^{1}P_{n_{i}}\otimes B_{n_{i}},
\end{equation}
with
\begin{equation}\label{B}
\nonumber  B_{n_{i}} = (-1)^{n_{i}+1}\sum_{l}\Bigg(g_{l}^{(i)}b^{(i)\dag}_{l}+g_{l}^{(i)*}b^{(i)}_{l}\Bigg).
\end{equation}
Let us assume $g_{l}^{(i)}\in\mathbb{R}$ for $i=1,2$ and all $l$.
This assumption does not restrict the problem. Dynamics of the model leading to pure dephasing can be solved in the interaction picture.
When working in this picture we put the sign  $\thicksim$ over the operators and coefficients.
Local interaction of the $i$th qubit with its own environment is described by unitary operator $\mathbf{\tilde{U}}_{i}(t)$.
The initially factorized total state, $\rho_{SE}(0)=\rho_{S}(0)\otimes\rho_{E}(0)$, evolves according to
$\mathbf{\tilde{U}}_{1}(t)\otimes \mathbf{\tilde{U}}_{2}(t)(\rho_{S}(0)\otimes\rho_{E}(0))\mathbf{\tilde{U}}_{1}^{\dag}(t)\otimes \mathbf{\tilde{U}}_{2}^{\dag}(t)$, where
$\mathbf{\tilde{U}}_{i}(t)=\sum_{n_{i}=0}^{1}P_{n_{i}}\otimes \tilde{u}_{n_{i}}(t)$ and
$\tilde{u}_{n_{i}}(t)= \textsf{T }e^{-i\int_{0}^{t}d\tau \tilde{B}_{n_{i}}(\tau)}$.
By partial tracing over the environment degrees of freedom, one finds the evolved system density matrix as
\begin{eqnarray}
\nonumber  \tilde{\rho}_{S}(t) &=& \sum_{n_{1},n_{2},n_{1}^{\prime},n_{2}^{\prime}} e^{\gamma_{n_{1},n_{2},n_{1}^{\prime},n_{2}^{\prime}}(t)} \\
\nonumber &\times& (P_{n_{1}}\otimes P_{n_{2}})\rho_{S}(0)(P_{n_{1}^{\prime}}\otimes P_{n_{2}^{\prime}}),
\end{eqnarray}
where
\begin{eqnarray}\label{C}
\nonumber e^{\gamma_{n_{1},n_{2},n_{1}^{\prime},n_{2}^{\prime}}(t)} &=&  Tr_{E}\{[\tilde{u}_{n_{1}^{\prime}}^{\dag}(t)\tilde{u}_{n_{1}}(t)\otimes\tilde{u}_{n_{2}^{\prime}}^{\dag}(t)\tilde{u}_{n_{2}}(t)]
\\ \nonumber &\times& \rho_{E}(0)\},
\end{eqnarray}
{which is the probability with which the environmental state transits to any other state.}
Note that the matrix $\tilde{C}(t)=(e^{\gamma_{n_{1},n_{2},n_{1}^{\prime},n_{2}^{\prime}}(t)})$ is semipositive definite and $\gamma_{n_{1},n_{2},n_{1},n_{2}}(t)=0$.
After some algebra one obtains
\begin{equation}\label{u}
\nonumber \tilde{u}_{n_{i}}(t)=\otimes_{l}e^{i\varphi_{i}(t)}\mathcal{D}[(-1)^{n_{i}+1}\beta_{l}^{(i)}],
\end{equation}
where $\mathcal{D}$ is the displacement operator, $\beta_{l}^{(i)}=g_{l}^{(i)}(1-e^{i\omega_{l}^{(i)}t})/\omega_{l}^{(i)}$ and the phase factor $\varphi_{i}(t)$ is given as
\begin{equation}\label{phase of u}
\nonumber \varphi_{i}(t)=\int_{0}^{t}dt_{1}\int_{0}^{t}dt_{2} \sum_{l}(g_{l}^{(i)})^{2}\sin(\omega_{l}^{(i)}(t_{1}-t_{2})).
\end{equation}
Regarding the environmental initial state as a tensor product of two-mode states, $\rho_{E}(0)=\bigotimes_{l}\rho_{l}^{12}$, one obtains   $\gamma_{n_{1},n_{2},n_{1}^{\prime},n_{2}^{\prime}}(t)=\Sigma_{l}\ln[\mathcal{X}_{l}^{n_{1}^{\prime},n_{2}^{\prime},n_{1},n_{2}}]$, where $\mathcal{X}_{l}^{n_{1}^{\prime},n_{2}^{\prime},n_{1},n_{2}}=Tr_{E}[(\mathcal{D}[2(n_{1}-n_{1}^{\prime})\beta_{l}^{(1)}]\otimes \mathcal{D}[2(n_{2}-n_{2}^{\prime})\beta_{l}^{(2)}] )\rho_{l}^{12} ]$ is the Wigner characteristic function of the state $\rho_{l}^{12}$ which can conveniently be determined for the Gaussian states.\\

For an initial state as $|\Psi_{12}\rangle=\frac{1}{\sqrt{2}}(|\textbf{\textsf{0}}\rangle+|\textbf{\textsf{1}}\rangle)$ $ $ $(|\textbf{\textsf{0}}\rangle=|01\rangle$ and $|\textbf{\textsf{1}}\rangle=|10\rangle)$, the evolved system density matrix in the interaction picture is
\begin{eqnarray}
\nonumber \tilde{\rho}_{s}(t) &=& \frac{1}{2}\Bigg(|\textbf{\textsf{0}}\rangle\langle\textbf{\textsf{0}}| +|\textbf{\textsf{1}}\rangle\langle\textbf{\textsf{1}}| \\
\nonumber &+& e^{\gamma_{0110}(t)}|\textbf{\textsf{0}}\rangle\langle\textbf{\textsf{1}}|+
e^{\gamma_{1001}(t)}|\textbf{\textsf{1}}\rangle\langle\textbf{\textsf{0}}|\Bigg),
\end{eqnarray}
where $\gamma_{0110}(t)=\gamma_{1001}(t)=\gamma(t)$. Finally, Eq. (\ref{Roh}) in the main text is given by
$e^{-iH_{\textit{S}}t}\tilde{\rho}_{s}(t)e^{iH_{\textit{S}}t}$ with $H_{\textit{S}} = H_{\textit{S}}^{(1)}+H_{\textit{S}}^{(2)}$.

\end{document}